\RequirePackage{fix-cm}
\documentclass[]{svjour3}
\usepackage{cite} 
\usepackage{enumitem}
\usepackage{algorithm} 
\usepackage{algpseudocode} 
\usepackage{wrapfig} 
\usepackage{graphicx}
\usepackage{url} 
\usepackage[hidelinks]{hyperref}
\usepackage{caption}
\usepackage{hyperref}
\usepackage[table]{xcolor}
\usepackage{titlesec}
\titleformat{\section}[block]{\bfseries\Large}{\thesection}{1em}{}
\titleformat{\subsection}[block]{\bfseries\large}{\thesubsection}{1em}{}
\titleformat{\subsubsection}[block]{\bfseries\normalsize}{\thesubsubsection}{1em}{}
 
\newcommand{\here}[1]{%
  \hypertarget{resp:#1}{}
  \fcolorbox{black}{black!15}{%
    \label{resp:#1}%
    \bfseries\scriptsize  {#1}%
  }%
}
\renewcommand{\here}[1]{}

\newcommand{\currentcolor}{black}

\newcommand{\setcolor}[1]{%
  \renewcommand{\currentcolor}{#1}%
  \color{#1}%
  \everypar{\color{\currentcolor}}%
}
\newcommand{\BLUE}{\setcolor{black}}
\newcommand{\BLACK}{\setcolor{black}}

\DeclareUnicodeCharacter{279C}{$\rightarrow$}
\begin{document}
\title{Shaky Structures: The Wobbly World of  Causal Graphs in Software Analytics}
\author{Jeremy Hulse \and Nasir U. Eisty \and
        Tim Menzies 
}

\institute{J.Hulse, T. Menzies \at
             Computer Science, NC State, USA \\ 
              \email{jphulse@ncsu.edu}   ,    
              \email{timm@ieee.org} 
           \and
           N. U. Eisty \at
           EECS, University of Tennessee Knoxville, USA \\
            \email{neisty@utk.edu}
}
\date{Received: date / Accepted: date}
\maketitle

\begin{abstract}
Causal graphs are widely used in software engineering to document and explore causal relationships.
Though widely used, they may also be wildly misleading.  
Causal structures generated from SE data can be highly variable.
This instability is so significant that conclusions drawn from one graph
may be totally reversed in another, even when both graphs are learned from the same or very similar project data. 

To document this problem,   this paper examines   causal graphs 
found by four  causal graph generators 
(PC, FCI, GES, and LiNGAM) when applied to 23  data sets, relating
to three different SE tasks:
(a)~learning how configuration options are selected for different  properties;
(b)~understanding how management choices affect software projects;
and (c)~defect prediction. Graphs were compared  between (a)~different projects exploring the same task; (b)~version $i$ and $i+1$ of a system;
(c)~different 90\% samples of the data; and (d)~small variations in the     causal graph generator.
Measured in terms of the Jaccard index of the number of edges shared by two different graphs, over half the edges 
were changed by these treatments. 

Hence, we conclude two things. Firstly, specific  conclusions found by causal graph generators about how two specific variables affect each other may not generalize since those conclusions could be reversed by minor changes in how those graphs are generated.
Secondly, before researchers can report supposedly general conclusions from causal graphs (e.g., ``long functions cause more defects''), they should test that such conclusions hold over the numerous causal graphs that might be generated from the same data.  


\end{abstract}

\section{Introduction}

Business users often demand explainable representations of their knowledge. For that purpose, some representations are deprecated  (e.g., neural nets, ensembles of decision trees) while others are more popular.  
For example, in 1973, David Lewis~\cite{Lewis1973} said, ``causation is something that makes a difference, and the difference it makes must be a difference from what would have happened without it''; i.e.
$T$ ``causes'' $Y$ if changing $T$ leads to a change in $Y$, keeping everything else constant. 

{\em Causal graph generators} are automatic algorithms that generate an acyclic graph supposedly reporting the causal connections within data. These are widely
used in software engineering; e.g., Nader et al.~\cite{Poshyvanyk24} use causal graphs to show programmers the significant relationships in their code data and conclude the best ways to effect change in that domain.  
This paper warns that:\vspace{1mm}

\BLUE 
\centerline{{\em Causal graph generators 
can be wildly misleading.}}\vspace{1mm}
\BLACK
\noindent Such generators must search all possible causal connections between all possible variables. This is an NP-hard task~\cite{PCParalell}, so, realistically,  causal graph generators must use heuristics
to find shortcuts in their search.
These heuristics lead to {\em causal
graph instability problem}; i.e., small changes to the input data of the generator can lead to very large changes in the generated causal graph.

The goal of this paper   is to document the   frequency and impact  of the SE causal graph instability. We will show:
   \begin{itemize}
       \item 
         Instabilities are rampant across multiple causal graph generators, multiple SE tasks,
          multiple SE data sets, and  multiple sub-samplings and tunings.
         \item   Our experiments explore some very  small changes to the training method. Yet even those small changes result in very different graphs.  
         \item Further, we document this impact and frequency in a repeatable and refutable
         manner. All our scripts and data are freely available online\footnote{ {\url{https://github.com/jphulse/Stability_Of_Causal_Graphs_Public}}}, thus allowing
         others to repeat/ refine/ or even refute our claims.
     \end{itemize}
The rest of this document is structured as follows.
   An initial motivation section (\S\ref{eg}) offers some    quick 
evidence that   demonstrates the existence of dauntingly large  causal graph instabilities. 
Next, our background section (\S\ref{back}) argues that such instabilities are a problem since causality is widely used  in SE.  
After that,
our experimental section (\S\ref{xp}) isolates the kinds of effects that can lead to causal graph instability. \BLUE
To show that these instabilities are a common effect, we look at the graphs that might be generated for a range of tasks   ({\em defect prediction}, {\em software configuration}, and {\em project management}) using a range of causal graph generators
 (PC, FCI, GES, LiNGAM~\cite{Spirtes2000,Spirtes2002,Chickering2002,Shimizu2006}).
Across this range of data sets and tasks,
we will observe very large causal graph instabilities: specifically,  our varying treatments, over 50\% to 80\% of the causal edges can change. \BLACK

 {\bf Important Caveat:} Note that this paper does not show that {\em all} causal graphs are unstable for {\em all} kinds of data analytics. That would be beyond the scope of any one paper. What we do show
is that for
 four
different causal graph generators (PC, FCI, GES, LiNGAM) applied to   23 data sets relating to three different SE tasks, such instabilities occur very frequently.

\section{Motivation}\label{eg}
This section offers the two experimental results that originally motivated us to write this paper.  Both examples
show surprisingly large instabilities in the graphs generated from causal graph generators.

 \begin{figure}
\begin{center}\small
{\bf Fig.~\ref{cause1}.a}: In this
causal graph from Ant Version 1.5,       ``bug''s,   cause ``loc'' (see left-hand-side).
\vspace{4mm}

\includegraphics[width=8cm]{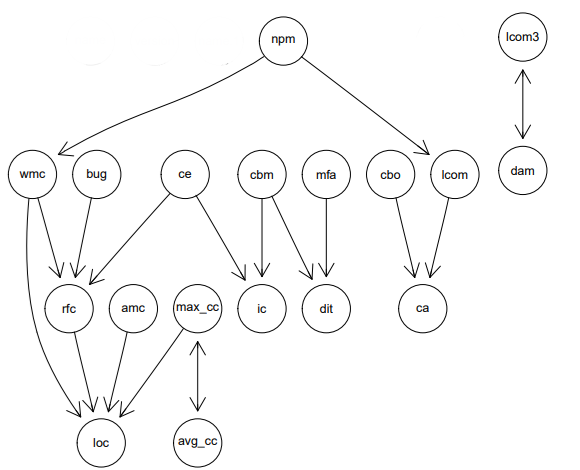}

\vspace{4mm}
 
{\bf Fig.~\ref{cause1}.b}: In this causal graph from  Ant Version 1.7, ``loc'' causes ``bug''s (see bottom-left).

\vspace{4mm}

\includegraphics[width=8cm]{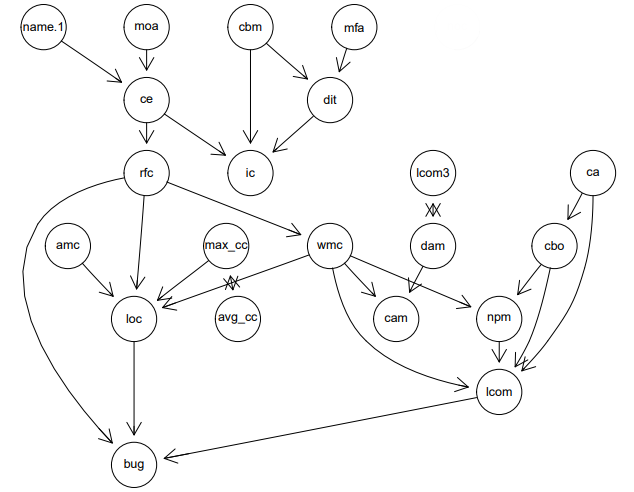}
\vspace{4mm}
\end{center}
\caption{Two (very different) causal graphs generated by the R pcalg package  (using \(\alpha = .01\)). For an explanation of the terms used in this figure, see Table~\ref{words}.}\label{cause1}
\end{figure} 

\begin{table}[!b]\scriptsize
\begin{tabular}{|l|p{10cm}|} 
\hline
\textbf{Attribute} & \textbf{Description} \\ \hline
name & The name of the software project \\ \hline
version & The version of the software project \\ \hline
name.1 & The extended name of the full software package (if one exists) \\ \hline
wmc & Weight Method Class or McCabe's complexity: Branch instructions in a class. \\ \hline
dit & Depth Inheritance Tree: Counts the number of ``fathers" a class has. \\ \hline
noc & Number of Children: Counts the number of immediate subclasses. \\ \hline
cbo & Coupling between objects: Counts the number of dependencies a class has. \\ \hline
rfc & Response for a Class: Counts unique method invocations in a class. \\ \hline
lcom & Lack of Cohesion of Methods: Calculates LCOM metric. \\ \hline
ca & Afferent Couplings that counts the  classes that depend on a given class. \\ \hline
ce & Efferent couplings: counts the number of classes that a given class depends on\\ \hline
npm & Number of Public methods: counts of the number of public methods. \\ \hline
lcom3 & LCOM* (Lack of Cohesion of Methods): Modified version of LCOM. \\ \hline
loc & Lines of code: Counts the lines of code, ignoring empty lines and comments. \\ \hline
dam & Data Access Metric is the ratio of the number of private attributes to the total number of attributes declared in the
class. \\ \hline
moa & Measure of Aggregation. Counts the number of data declarations (like class-level variables) whose types are user-defined classes.\\ \hline
mfa & Measure of Functional Abstraction. Ratio of inherited methods to all methods. \\ \hline
cam & Cohesion Among Methods of Class that measures the similarity of parameter lists of methods in a class, indicating method cohesion.\\ \hline
ic & Inheritance Coupling that counts the number of parent classes to which a class is coupled. It indicates the class's dependency on inherited methods.\\ \hline
cbm & Coupling Between Methods - measures total complexity of coupling between class methods.\\ \hline
amc & Average Method Complexity -   the average complexity of methods in a class. \\ \hline
max\_cc & Maximum Cyclomatic Complexity -   max. cyclomatic complexity among the methods of a class. \\ \hline
avg\_cc & Average Cyclomatic Complexity- calculation of the average cyclomatic complexity of methods in a class. \\ \hline
bug & Number of bugs \\ \hline
\end{tabular}
\caption{Variables see in Fig.~\ref{cause1}.}\label{words}
\end{table}

\subsection{ Example 1: Instabilities between two versions}

In this first example, we keep the causal graph {\em generator   constant} but make a small
{\em change to the data}.

Specifically, we apply the PC causal graph generator (described later in this paper)  to two   versions of the same defect data sets. The nodes in these graphs use the attributed defined in Table~\ref{words}.
The resulting graphs are shown in Fig.~\ref{cause1}. 

 Fig.~\ref{cause1}.a and Fig.~\ref{cause1}.b show many differences  in the causal graphs from two different releases of the same software project:
 \begin{itemize}
     \item 
  Fig.~\ref{cause1}.b tells us that {\em loc} (lines of code) causes {\em bugs}. This seems to be a reasonable conclusion (justification:   the more lines of code, the more likely a programmer will make a mistake).
  \item  
      But  Fig.~\ref{cause1}.a, which is from an earlier version of the same project, offers the reversed causal connection. In this graph, {\em bugs} cause changes to {\em rfc} (how many methods wake up when a message arrives at a class), which, in turn, causes changes to {\em loc}. 
\end{itemize}
To say that another way, in one release, lines of code seem to {\em cause} bugs
 but in another, lines of code are the {\em consequence} of bugs (this is such a  very strange result that the conclusion
 of this paper will be to not trust  this data).

  \BLUE
\subsection{Example 2: Instabilities between two generators} 

\here{R1.1} 
In this second example, we keep the
{\em data constant}   but make changes to the
{\em causal graph generation} process.

\begin{wrapfigure}{r}{2.25in}
\includegraphics[width=2.25in]{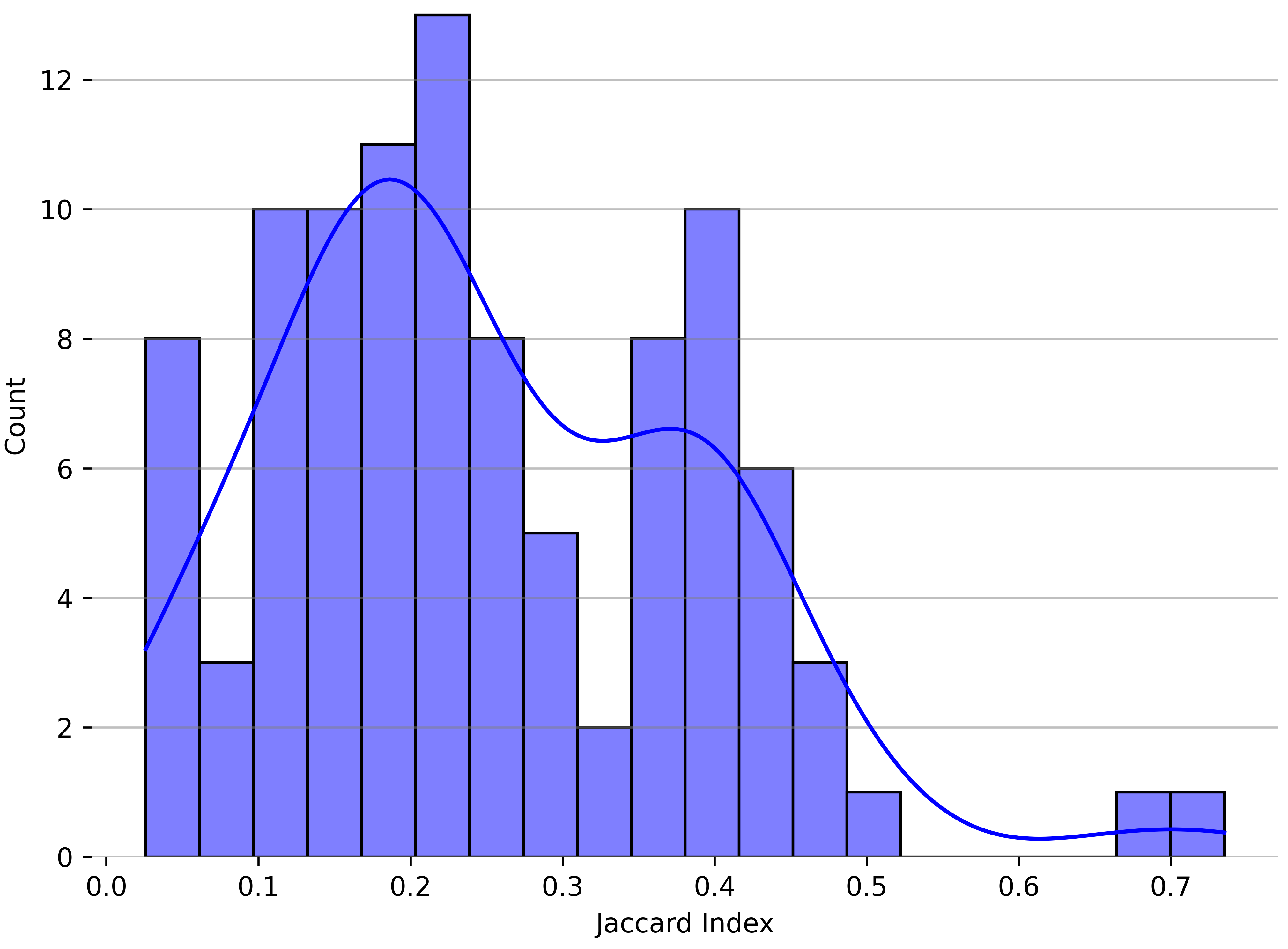}
\caption{Jaccard index results seen in 100 experiments where two causal graph generators are applied to the same data set. }\label{overlap}
\end{wrapfigure}
Specifically, this section explores 
23 data sets from the domains
of 
{\em defect prediction}, 
{\em software configuration}, and 
{\em project management}.  The defect prediction data has the structure shown in Table~\ref{words}, and the other data sets are described later in this paper (see \S\ref{Xdata}).
Within this data, 
 100 times:
 \begin{itemize}
 \item
 We select at random one data set 
 \item We select at random,  two causal graph generators from the set   \{PC, FCI, GES, LiNGAM\} (these causal graph generators are described later in this paper).
 \item
These generators are then applied to the data (using their default settings).
\item
We report the overlap in the causal connections found by this pair of generators.
 \end{itemize}
\here{R1.3} 
For  graphs with unweighted directed edges,
Zweig~\cite{zweig2016network} reports that there are many ways to measure graph overlap, such as the Jaccard index:
\begin{itemize}
\item A causal graph has  nodes $N$ and edges $E$. 
\item Edges are the same if they  
start and end on the same node (so by this definition,   edges using the   same nodes are different if they have a different direction).
\end{itemize}
The Jaccard index for two graphs divides the  
     number of shared edges  by the total number of edges in both graphs:

    \begin{equation}
    J=\frac{|E_1 \cap E_2|}{|E_1 \cup E_2|}
     \end{equation} 

     This Jaccard index has the range $0 \le J \le 1$, where zero denotes no overlap and one denotes perfect overlap. We say that {\em larger} $J$ values indicate {\em more} stability.

Fig.~\ref{overlap} shows the observed Jaccard indexes. If two generators always found  the same edges on the same data,   then   these results would have a Jaccard index of 1.0. However, in Fig.~\ref{overlap}, the median Jaccard is 0.25.

To say that another way, for the same data set, three-quarters of the edges found by one generator are usually not found by another.

\subsection{A Comment on these Examples}

  By themselves,   these examples
are not enough to  make general conclusions:
\begin{itemize}
\item
\here{R2.2}
Example 1 might just be a quirk of
data collection within one specific project.
\item
The random nature of Example 2 might make it overstate the problem. For example, when we restrict the data set selection to back-to-back releases of software from the same project (as done in 
\S\ref{qrq1}), we get much higher stability. Hence, we offer Example 2 as a motivator but not definitive proof. 
\end{itemize}
What these examples demonstrate is that causal graph generators can produce wildly different outputs. 
To investigate and document this effect,   the rest of this paper performs controlled experiments that explore specific classes of changes. 

\BLACK

\section{Background } \label{back}
\subsection{Causality}
 Proponents of causality argue that it is a more informative concept than mere correlation. If we
 only say that two variables X and Y are correlated, then this leaves open  many questions:  
 \begin{itemize}
     \item Does X cause Y?
     \item Does Y cause X?
     \item Is there some third hidden confounding variable that actually 
     is the cause of X and Y?
 \end{itemize}
 On the other hand, when we can say X causes Y, we can move from merely observing the world to actually changing it. Nadir et al. call this ``climbing Pearl's Ladder of Causation''~\cite{Poshyvanyk24}.  Judea Pearl was an early pioneer in causality in AI.  In his seminal textbook with  Mackenzie~\cite{pearl2018book},  Pearl said that causal inference explores   questions of association (what is?),
counterfactual interventions (what if?), and pure counterfactuals (why?). 

Different levels of human cognitive styles, says Pearl, match different levels of this hierarchy. For example, merely reporting ``what is?'' via correlation
(or some method that infers $y=f(x)$) is level one.
Deciding ``what to do?''  (i.e., what interventions to take) is a higher-level task that is more cognitively difficult. This requires questions about ``what if?'' and ``what if I do X?''.  
Finally, above  the ``doing'' level is the ``imagining'' level that explores
issues of ``why?'' and ``was it X that caused Y?'' and ``what if I had acted differently?''. 

\subsection{Causal Graphs}
Causal graphs are a data structure to showcase causal relationships between a set of random variables.
 Causal graphs, also known as causal diagrams (and perhaps causal Bayesian networks), are a graphical representation used to illustrate and analyze the causal relationships between 
 \BLUE variables in a system~\cite{Pearl1995,pearl2009causality}.  \here{R2.3} \BLACK

In causal graphs, nodes represent variables or events, while directed edges (arrows) indicate causal influences from one variable to another. 
Causal graphs are typically structured as Directed Acyclic Graphs (DAGs), meaning they have directed edges and no cycles, ensuring a well-defined ordering of events or variables~\cite{williams2018directed}.
Fig.~\ref{cause1} shows two examples of such graphs.


\begin{table}
\caption{  Kinds of causal analysis seen in SE. 
 As discussed here, none of these approaches are guaranteed to produce correct models.  }\label{kinds}
\footnotesize
\begin{tabular}{|p{.95\linewidth}|}\hline
\BLUE 
 All the experiments of this paper are  {\bf generative studies} that  start without preexisting assumptions.
 Causal graphs are then generated via automatic tools, then  examined for insights. For example, 
 Johnson et al.~\cite{ 10.1145/3377811.3380377} 
 explore the edges in automatically generated causal graphs, looking for the root cause of a test failure.

 While useful in some test domains, generative studies   can yield absurd conclusion  such as (a)~the Fig.~\ref{cause1}  conclusion that bugs cause lines of code; or (b)~spurious correlations like the (in)famous correlation between chocolate consumption  per 10 million persons and Nobel laureates per country (in their highly satirical article,
Messerli~\cite{messerli2012chocolate} reports that chocolate consumption correlates to   Nobel prizes at $r=0.79, P<0.0001$).   
\\\hline \BLUE
 \here{R2.1b}
  {\bf Confirmatory studies} begin with established assumptions (expressed as an initial set of hand-drawn causal graphs), which are then tested by generating a causal graph and checking which prior assumptions
  are found in the data.
  
  While a useful prudence check on intuitions, relying solely on confirmatory analysis have two problems:
  \begin{itemize}
  \item {\em Overlooking  novel insights:} 
  Breiman's famous ``Two Cultures'' paper~\cite{breiman2001statistical} highlights the limitations of restricting analysis to expected outcomes.
  The Wikipedia page ``List of discoveries influenced by chance circumstances''   includes pacemakers, microwave ovens, vulcanized rubber, penicillin,     Big Bang radiation, the Michelson-Morley effect, Viagra, saccharin, etc. 
  \item {\em Over-reliance on analyst bias:} Manually drawn causal graphs may not be a reliable ground truth. For example, we recently had a paper rejected for comparing causal graphs from algorithms with those drawn by SE Ph.D. students
  (reviewers complained  that  manually drawn graphs were not a trustworthy source). 
\end{itemize} 

 \\\hline \BLUE
 \begin{wrapfigure}{r}{1.5in}
\includegraphics[width=1.5in]{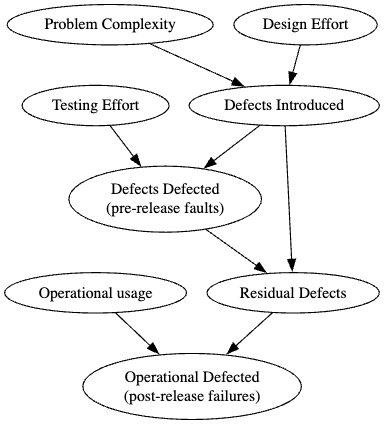}
\end{wrapfigure} Confirmatory studies may be followed by
{\bf Post-patch}  studies that adjust  causal models based on unexpected findings. 
For example, reliability engineers~\cite{musa09}  
propose
shipping software once the expected number of pre-release
faults fall below a certain threshold.
 This approach assumes the causal model:
\[
\mathit{pre\mbox{-}release\;faults} \rightarrow \mathit{post\mbox{-}release\; failures}
\]
Using data from real-world projects, 
Fenton \& Neil~\cite{fenton1999critique,Fenton2000SoftwareMR}   report many examples of few pre-release faults but
many post-release failures. To reconcile this anomaly, they expand the above simple causal model 
to incorporate factors like system usage, defect detection, and design effort. This revised model (shown at right)
shows how (e.g.) systems with no testing could exhibit few faults but many failures. Even though this extended causal model lacked empirical validation, it was still accepted by reviewers at senior SE forums~\cite{fenton1999critique,Fenton2000SoftwareMR}.

\vspace{1mm}
Such post-patching runs the risk of shoddy science, of ``making it up as you go along''. However, as Ernst  \& Baldassarre~\cite{Ernst2023} remark, patching is valid if it adheres to some revision process that was pre-specified before the data is analyzed. 
 \BLACK\\\hline
\end{tabular}
\end{table}

\newpage
\BLUE
 \here{R2.1a}{ In general, as seen in Table~\ref{kinds}, these graphs can be used in three many  ways  including:}
\begin{enumerate}
  \item
  {\em Confirmatory studies}: checking a   pre-experimental belief; 
  \item
  {\em Generative studies}: generating a new belief, from data;
  \item
  and {\em Post-patching}: adjusting a graph to fit the data.
  \end{enumerate}
\here{AE.2c} A key point to make here is that 
 all these kinds of studies  rely in various ways to 
  the structures generated by automatic causal graph generators. 
If quirks in graph generation can lead to idiosyncratically different graphs, then:
\begin{itemize}
\item
Confirmatory studies can be misleading if the learned causal graph misses the effect being explored by the analyst. 
\item
Exploratory studies can lead to incorrect conclusion when the graph generator randomly makes up, or omits, some causal link.
\item
Post-patching can waste much effort while exploring   suspiciously generated causal connections. 
\end{itemize}
Hence, we are concerned that  
causal structure instability can   negatively impact many different kinds of causal analysis. 
 
   \BLACK

\begin{table}[t!]
 
    \caption{ Applications of causality in SE. From~\cite{USES}.}\label{field}
\BLUE

  \begin{center}
  \begin{minipage}{2in}  \scriptsize 
  \begin{tabular}{ll}  
        \textbf{Software Engineering Fields} &  papers  \\
        \hline
        Testing & 17 \\
        Deployment & 9 \\
        Design & 7 \\
        Development & 5 \\
        Maintenance & 5 \\
        Collaboration & 1  
    \end{tabular}~\\~\\~\\~\\
\end{minipage}\hspace{0.25in}\begin{minipage}{2in}  \scriptsize
  
  \begin{tabular}{ll}  
        \textbf{Causal Reasoning Applications} &  papers \\
        \hline
        Causality Graph & 17 \\
        Groundwork to Facilitate Causality & 12 \\
        Bayesian Model & 5 \\
        Granger Causality Test & 3 \\
        Difference-in-Differences Model & 2 \\
        Counterfactual Prediction & 2 \\
        Statistical Analysis & 1 \\
        Evidential Network & 1 \\
        Propensity Model & 1  
    \end{tabular}
  
    \end{minipage}
    \end{center}
    \BLACK
\end{table}



 \subsection{ Causal Graphs in SE}

Our reading of the SE literature is that causal graphs are the most widely used causal reasoning application~\cite{USES}
(and those graphs may be used in a confirmatory or generative or
post-patch manner).
\BLUE
For example,
  Table~\ref{field}  shows results from a recent survey on the use of causality in SE. \BLACK
The rest of 
this section offers notes on some of the ways
causality is used in SE (and the following   is just a sample,
not a complete list of applications).

 When software fails, engineers might find causal graphs useful to map out the sequence of events leading up to the failure. 
For example, if an application crashes under certain conditions, a causal graph could help trace back through the code and system states to identify which specific inputs and interactions caused the crash. 
This makes it easier to pinpoint and fix the root cause of the issue~\cite{10.1145/3635709, 10.1145/3377811.3380377}.

Causal graphs are also valuable in performance optimization~\cite{10.1007/978-3-030-59152-6_19, 10.1145/3492321.3519575, wu2019employing}. 
Engineers can use them to identify bottlenecks and understand how different parts of a system interact to affect overall performance. 
For instance, if a web application is slow, a causal graph can help visualize the causal relationships between user actions, server requests, database queries, and response times. 
By understanding these interactions, engineers can make targeted optimizations, such as improving database indexing or optimizing server-side code, to enhance performance.

Another  use of causal graphs in software engineering is in change impact and risk analysis~\cite{HU2013439}. 
Before making modifications to a codebase, engineers can create causal graphs to predict how changes might propagate through the system. 
For example, when adding a new feature, a causal graph can help identify which existing modules and functionalities will be affected, ensuring that potential issues are addressed before they arise. 
This helps in better planning and risk management, reducing the likelihood of introducing new bugs or performance issues.

Effective testing and verification could also use  causality~\cite{10.1145/3635709, 10.1145/3377811.3380377}. 
Testing involves comprehending how changes in the codebase affect the system's behavior. 
By understanding causality, engineers could design tests that accurately capture the effects of code modifications, ensuring new features function correctly and existing functionality remains intact.

In multi-threaded or distributed systems, 
concepts of causality play a large role in 
discussions about managing concurrency and synchronization~\cite{485846, 6848128}. 
Understanding the order and dependencies of events is very useful for preventing issues like race conditions, deadlocks, and data inconsistencies, which can arise when causal relationships between concurrent operations are not properly managed. Similarly, performance optimization often hinges on 
some understanding of how one event might cause another~\cite{10.1007/978-3-030-59152-6_19, 10.1145/3492321.3519575, wu2019employing}. 
By understanding how interactions between components affect performance, engineers can make targeted optimizations to improve overall system efficiency.

Maintaining traceability in software development involves tracking the origins of requirements, changes, and their implementation. 
This process can be defined as understanding the causal relationships between various development artifacts, ensuring changes are properly documented and their impacts are understood~\cite{6644251}.

\BLUE
\here{R1.2c} 
Another use instance of causal graphs in the SE literature is in the generation of metamorphic test cases. Clark et al.~\cite{clark2023metamorphic}  generate random causal graphs and then make synthetic programs that match the generated graph.  Then, after this, they are able to generate metamorphic tests to evaluate that code based on their generated causal graph.

There have been a handful of literature reviews conducted into the SE literature for causal reasoning, being the broader field in which causal inference and causal discovery lie.  Two such reviews were conducted by Siebert and Giamattei ~\cite{SIEBERT2023107198}~\cite{giamattei2024causal}.  In both of their reviews, it is noted that the current field of interest, particularly in software engineering, is very fragmented, which makes their work as literature reviews or broader examinations extremely valuable. Some examples of usage of causal reasoning in software engineering from Giamattei et al.~\cite{giamattei2024causal} include:
\begin{itemize}
    \item Fault localization
    \item Testing and analysis
    \item KPI prediction
    \item Defect prediction
    \item Anomaly detection
    \item Threats modeling
\end{itemize}
Further still, Giamattei et al.~\cite{giamattei2024causal} detail observed usage rates of various causal discovery algorithms.  They observed that Granger causality~\cite{granger1969investigating} is used most frequently, followed by custom algorithms, then followed closely by PC~\cite{Spirtes2000}. Several of the custom algorithms are also modifications to the PC algorithm.  FCI, GES, and LiNGAM are also observed in their literature review~\cite{Spirtes2002,Chickering2002,Shimizu2006}.

\BLACK

Finally, it is argued that systems designed with a clear understanding of causality tend to be more reliable and maintainable~\cite{7321183}. 
By comprehending how different parts of the system interact and affect each other, engineers can build more robust systems that are less prone to errors and easier to extend or modify.


\subsection{Significance of Instabilities in Causal Graphs}

Section~\ref{eg}
offered specific examples of causal graph instability.
In this section, we ask what the effects of these instabilities might be.

For software engineering applications,
instabilities in causal graphs can significantly impact the accuracy, reliability, and effectiveness of causal reasoning in software engineering~\cite{chindelevitch2012assessing}. 
Here are several reasons why addressing these instabilities is crucial 

 \textbf{Accuracy of Analysis.} Causal graphs are used to make predictions and inferences about the relationships between variables~\cite{liang2021normalized}. If the graph is unstable due to outdated or incorrect causal relationships, the resulting analysis can be inaccurate. For example, a performance optimization based on an outdated causal graph may fail to address the actual bottlenecks in a system, leading to suboptimal improvements or even performance degradation~\cite{10.1007/978-3-030-59152-6_19, 10.1145/3492321.3519575}.



  \textbf{Scalability and Maintainability.} Software systems need to scale and evolve over time to meet changing requirements and increasing user demands. Instabilities in causal graphs can hinder this process by making it difficult to predict the impact of changes accurately~\cite{9218193}. This can lead to increased maintenance efforts as engineers struggle to diagnose and fix issues that arise from incorrect causal assumptions~\cite{7321183}. In a large-scale system, such instabilities can propagate, causing widespread problems that are costly and time-consuming to address.

  \textbf{Data-Driven Decision Making.} In data-driven environments, decisions are often based on the insights derived from causal analysis. If the causal graph is unstable, the insights and recommendations generated from it can be misleading~\cite{10.1145/3636423}. This can lead to poor decision-making, whether it's in optimizing business processes, improving user experiences, or implementing new technologies. For example, in a recommendation system, inaccurate causal relationships can lead to irrelevant or ineffective recommendations, reducing user satisfaction and engagement.

\section{Experiments}\label{xp}

\subsection{Research Questions}
Our experiments address      five research questions:
\begin{itemize}
    \item 
   {\bf RQ1:} {\em  Do the graphs  demonstrate intra-project stability across different
    releases? }
    For    SE data,  various releases,  we can   explore
how causal graphs change within the lifetime of one project.
   In that corpus, over half the causal effects found in release $i$ are missing in release $i+1$.
    \item   
    {\bf RQ2:} {\em Do the graphs demonstrate inter-project stability across different projects? }
       In our corpus, over two-thirds of the causal connections present in one project
       may be different in another.  
      \item  
    {\bf RQ3:} {\em Does parameter tuning affect stability?} In our corpus, seemingly innocuous changes to tuning parameters can generate radically different causal graphs.
       \item 
     {\bf RQ4:}  {\em  How much do causal graphs change if we remove only small fractions
     of the training data?} Removing as little as 10\% of training data can lead to very
     different graphs.  
     \item 
 {\BLUE} {\bf RQ5:}  {\em  Do the instabilities reported above hold for other causal
generators and other tasks?} 
     For different tasks, we find different levels on instability. But for all tasks, we found large numbers of instabilities.    
\end{itemize} 
At first, we explore {\bf RQ1, RQ2, RQ3, and RQ4} for defect prediction and the PC causal graph generator. Then, for {\bf RQ5}, we repeated that for different causal graph generators 
(FCI, GES, LiNGAM) and for a broader range of problems (defect prediction
{\em and} software configuration problems {\em and}  project management problems). What will be seen is that the effects seen for PC+defect prediction can also be seen when the other generators are applied to other tasks.  Rather than fill this paper with graphs that offer no new information, 
{\bf RQ5} will instead offer a summary of all the effects across all the tasks. {\BLACK}


\subsubsection{Data sets}\label{Xdata}
Initially, we explore six {\bf defect prediction data} sets from widely used open-source JAVA projects.
These projects contain static attributes that describe thousands of classes (using the attributes listed in   Table~\ref{words}). The datasets used are from Jureczko et al.~\cite{defectDatasets} and are versioned as follows:
\begin{itemize}
    \item  Ant versions 1.3-1.7, 
    \item Camel versions 1.0, 1.2, 1.4, \& 1.6, 
    \item Ivy versions 1.1, 1.4, \& 2.0, 
    \item Prop releases 1 - 6, 
    \item Velocity versions 1.4 - 1.6, 
    \item Xerces versions 1.2 - 1.4. 
\end{itemize} 
\BLUE \here{AE.3} We also address two other SE tasks.
\begin{itemize}
\item {\bf Configuration data:}
{\em SS-Models (SS-M, SS-P)}:
These data sets were obtained from the software configuration literature~\cite{peng2023veer}. The data was collected by running different software projects configured in different ways (selected at random) and then collecting different performance metrics (runtimes, CPU usage, etc.). The goal of these datasets is to find a configuration of the software that best optimizes the overall software goals for each specific project.

\item {\bf Process Data:}
{\em XOMO\_Flight,
XOMO\_Ground, 
XOMO\_OSP}
were  introduced by   Menzies et al.~\cite{menzies2005xomo} and are a general framework for Monte Carlo simulations that combine four COCOMO-like software process models from Boehm's group at the University of Southern California. The overall goals for XOMO are to:
\begin{itemize}
    \item Reduce risk;
    \item Reduce effort;
    \item Reduce defects;
    \item Reduce development time.
\end{itemize}
The available XOMO models here all come from NASA's Jet Propulsion Laboratory (JPL).   Flight and Ground are general descriptions of all JPL's flight and ground software, and OSP and OSP2 are two versions of the flight guidance system of the Orbital Space Plane. In terms of complexity, we know that nasa93dem is the simplest of the five, followed by OSP and OSP2, which are similar, then Ground is a bit more complex, and finally, Flight is the most complex of all. 
\end{itemize}
All these data sets are summarized in Table~\ref{datas}. At first glance,
there seems to be far more defect data than anything, but this is due to
the fact that for the defect data, we have information
on multiple releases of those five software projects. \BLACK
 
   
\begin{table}[!t]
\centering
\caption{Revision datasets and attributes.}\label{datas}
\label{Revision_Dataset_Attributes}
\BLUE
\begin{tabular}{|l|l|r|r|r|r|}
\hline
   &  &  & \multicolumn{1}{l|}{\textbf{Independent  }} & \multicolumn{1}{l|}{\textbf{Dependent  }} & \multicolumn{1}{l|}{\textbf{Percent  }} \\
\textbf{Group}   & \textbf{Dataset} & \multicolumn{1}{l|}{\textbf{Rows}} & \multicolumn{1}{l|}{\textbf{  Variables}} & \multicolumn{1}{l|}{\textbf{  Variables}} & \multicolumn{1}{l|}{\textbf{  Defective}} \\ \hline
Config          & SS-N          & 53662                              & 17                                                  & 2                                                  & n.a.                                                  \\  
           & SS-P          & 1023                               & 11                                                  & 2                                                  & n.a.                                                  \\ \hline
          & XOMO\_Flight   & 10000                              & 27                                                  & 4                                                  & n.a.                                                  \\  
Process         & XOMO\_Ground   & 10000                              & 27                                                  & 4                                                  & n.a.                                                  \\  
          & XOMO\_OSP      & 10000                              & 27                                                  & 4                                                  & n.a.                                                  \\ \hline
         & Ant-1.3       & 125                                & 20                                                  & 1                                                  & 16                                                   \\  
           & Ant-1.4       & 178                                & 20                                                  & 1                                                  & 22                                                   \\  
           & Ant-1.5       & 293                                & 20                                                  & 1                                                  & 11                                                   \\  
           & Ant-1.6       & 351                                & 20                                                  & 1                                                  & 26                                                   \\  
           & Ant-1.7       & 745                                & 20                                                  & 1                                                  & 22                                                   \\  
           & Camel-1.0     & 339                                & 20                                                  & 1                                                  & 4                                                    \\   
           & Camel-1.2     & 608                                & 20                                                  & 1                                                  & 36                                                   \\  
           & Camel-1.4     & 872                                & 20                                                  & 1                                                  & 17                                                   \\  
Defect          & Camel-1.6     & 965                                & 20                                                  & 1                                                  & 19                                                   \\  
           & Ivy-1.1       & 111                                & 20                                                  & 1                                                  & 57                                                   \\  
           & Ivy-1.4       & 241                                & 20                                                  & 1                                                  & 7                                                    \\   
           & Ivy-2.0       & 352                                & 20                                                  & 1                                                  & 11                                                   \\  
           & Synapse-1.0   & 157                                & 20                                                  & 1                                                  & 10                                                   \\ 
           & Synapse-1.1   & 222                                & 20                                                  & 1                                                  & 27                                                   \\  
           & Synapse-1.2   & 256                                & 20                                                  & 1                                                  & 34                                                   \\  
           & Xerces-1.2    & 440                                & 20                                                  & 1                                                  & 16                                                   \\  
           & Xerces-1.3    & 453                                & 20                                                  & 1                                                  & 15                                                   \\  
           & Xerces-1.4    & 588                                & 20                                                  & 1                                                  & 74                                                   \\ \hline
\end{tabular}\BLACK
\end{table}

    \subsection{Algorithms}\label{algorithms}
\here{AE.5} This study passes all the the data of Table~\ref{datas} through four different causal graph generators.

  Le et.al.~\cite{PCParalell}  present two methods of learning and making causal structures:
 \begin{itemize}
 \item
The first method they list is {\em search and score}; in this technique, the program will cycle through and score all possible directed acyclic graphs connecting $N$ variables in the data,  keeping the best one(s) (where ``best'' is based on a scoring function defined by the developer). They also mention that this approach is largely infeasible as it is NP-Hard to learn all networks in this way~\cite{PCParalell}.  Accordingly,
 practical causal graph generators have to rely on some second method.
      \item 
 The second method of learning these structures is more heuristic. For example,
 a conditional independence test is used in order to remove any relationships identified as conditionally independent and, therefore, not causal. 
 While this approach is more computationally feasible and scalable,
 it does miss certain causal relationships. Also, depending on certain
 idiosyncratic decisions (on the ordering within which variables are searched,
 what thresholds are used to restrain the search, and how to break ties between
 similar influences), these heuristics can lead to very different causal
 structures.     Using this second method, many  causal graph generators have been proposed, including   
 (PC, FCI, GES, LiNGAM~\cite{Spirtes2000,Spirtes2002,Chickering2002,Shimizu2006}).
  \end{itemize}
Having worked in this area,
we can assert that finding a causal graph 
generator can be a challenging task.
One reason for this is methodological-- 
researchers in this arena rarely
share resources.  
Chadbourne and Eisty~\cite{USES} explained in their systematic literature review that although the exploration of a significant number of papers on software engineering and causal reasoning was enlightening, it was challenging to draw strongly supported conclusions about the applications of individual causal reasoning techniques. 
This difficulty arose due to the minimal number of papers that explicitly utilized these techniques. 
This scarcity can hinder progress and innovation in the field. 

\begin{table}[!t]
\caption{Assumptions of different causal graph generators.}\label{particular}
  \here{AE.2b} \centering\scriptsize
   \BLUE
  \begin{tabular}{|p{.7in}|p{.6in}|p{.6in}|p{1in}|p{1in}|}
        \hline
        & PC \cite{Spirtes2000} & FCI \cite{Spirtes2002} & GES \cite{Chickering2002} & LiNGAM \cite{Shimizu2006} \\ \hline
        Authors & Spirtes et al. \cite{Spirtes2000}  & Sprites \& Zhang \cite{Spirtes2002} & Chickering \cite{Chickering2002} & Shimizu et al. \cite{Shimizu2006} \\ \hline
        Faithfulness assumption required? & Yes & Yes & Some weaker condition required (not totally clear yet) & No \\ \hline
        Specific \newline assumptions on data \newline distributions required? & No & No & Yes (usually assumes linear-Gaussian models or multinomial distributions) & Yes \\ \hline
        Mechanisms to handle confounders? & No & Yes & No & No \\ \hline
    Linearity \newline assumptions? & not inherent & not inherent & usually linear & definitely linear\\\hline
        Links to \newline working code &  \href{https://cran.r-project.org/web/packages/pcalg/index.html}{pcalg}~\cite{PCALG1};   
           \href{https://causal-learn.readthedocs.io/en/latest/search_methods_index/Constraint-based%20causal%20discovery%20methods/PC.html}{causal learn PC alg}~\cite{zheng2024causal}
          &  
             \href{https://cran.r-project.org/web/packages/pcalg/index.html}{pcalg}~\cite{PCALG1};    
             \href{https://causal-learn.readthedocs.io/en/latest/search_methods_index/Constraint-based%20causal%20discovery%20methods/FCI.html}{causal learn FCI alg}~\cite{zheng2024causal}
        &  \href{https://cran.r-project.org/web/packages/pcalg/index.html}{pcalg}~\cite{PCALG1};      \href{https://causal-learn.readthedocs.io/en/latest/search_methods_index/Score-based%20causal%20discovery%20methods/GES.html}{causal learn GES alg}~\cite{zheng2024causal}
       &  \href{https://causal-learn.readthedocs.io/en/latest/search_methods_index/Causal%20discovery%20methods%20based%20on%20constrained%20functional%20causal%20models/lingam.html}{causal learn LiNGAM algs}~\cite{zheng2024causal} 
        \\ \hline
  \end{tabular}
   \BLACK
\end{table}

\BLUE
Table~\ref{particular} shows some of the causal graph generators currently accessible. 

\here{AE.2a} Note that different causal graph generators make different assumptions about the data,
and handle special cases in different ways. 
 \begin{enumerate}
    \item  The \textbf{PC}  algorithm takes a constraint-based approach for causal discovery in which conditional independence is tested repeatedly and iteratively to construct a DAG of connected edges that represent causal relations ~\cite{Spirtes2000}.
    \item The \textbf{FCI} algorithm is a modification of the PC algorithm that utilizes the conditional independence tests from the PC algorithm along with additional constraints in order to develop causal relations. FCI was designed to work in the presence of latent confounders\footnote{Causal confounders are variables that influence both the independent variable (the presumed cause) and the dependent variable (the presumed effect), causing a spurious association between the two~\cite{spirtes2013causal}.}
    \item The \textbf{GES}  algorithm is a score-based causal discovery method in which some scoring criteria are used to adjust edges greedily until a local extrema or other criteria are met by the evaluation of the graph~\cite{Chickering2002}.
    \item The \textbf{LiNGAM} (Linear Non-Gaussian Acyclic Model) algorithm is a different category of causal discovery described on causal-learn as a ``causal discovery method based on constrained functional causal models''~\cite{zheng2024causal}. LiNGAM operates by assuming that the variables in the dataset are non-Gaussian and attempts to exploit the nature of the data by performing   ICA (independent component analysis).
    This is a method to  identify the independent components of the data, which are assumed to correspond to the underlying causal variables
    ~\cite{Shimizu2006}.
\end{enumerate}
We focus on these  four algorithms since they
are publicly available
(in the ``R'' systems and elsewhere~\cite{zheng2024causal})
and represent a range of algorithmic methods. For example, 
our algorithms are constraint-based, score-based, and constrained causal model-based.  
Also, they make different assumptions about the data.
\here{R1.2a} For example,  in the context of causal reasoning,
{\em sufficiency}   is the assumption that algorithms can access all the relevant variables that could be influencing the relationships you are studying.
\begin{itemize}
\item
PC (and some other causal discovery algorithms) makes a sufficiency assumption. Hence, if there are unmeasured confounders, they may produce incorrect causal structures (since conditional independencies can be misleading when latent variables are present).
\item
However, the FCI algorithm is designed to relax the sufficiency assumption. It can still provide valid causal inferences even when there are unmeasured confounders. It outputs a partial ancestral graph (PAG), which can represent possible causal relationships in the presence of latent variables. The PAG will include special markings to indicate the possible existence and location of latent confounders.
\end{itemize}
 For another example of how these algorithms differ, consider the {\em faithfulness} assumption (also known as stability or the causal Markov condition)  that assumes if two variables appear independent given a third variable, it's because the causal graph truly implies that independence, and not because some coincidental combination of parameter values is masking a real dependence.
 
PC  and FCI assume faithfulness.  Hence, if there are unmeasured confounders, they may produce incorrect causal structures (since conditional independencies can be misleading when latent variables are present).
But LiNGAM relies on different assumptions.  This algorithm uses the functional form of relationships,  rather than conditional independencies, to infer causality, so it is not directly affected by violations of faithfulness.
 
\here{R1.2b}
As to the linearity of effects between variables,
these algorithms take different approaches:
\begin{itemize}
\item
PC and   FCI are constraint-based algorithms that rely on conditional independence tests, which can be implemented under various statistical assumptions. While implementations often assume linear Gaussian relationships (e.g., via partial correlation tests), the algorithms themselves are not inherently limited to linear models; nonparametric or kernel-based tests can also be used. However, such extensions can be more computationally intensive and less statistically powerful with limited data.
\item
  GES is a score-based approach that typically assumes linearity and Gaussian noise when using BIC or related scoring criteria. While GES can, in principle, be extended to non-linear settings using alternative scoring functions, the standard formulation is linear-Gaussian.
\item
 Unlike the others, LiNGAM explicitly assumes a linear causal model with non-Gaussian noise. This assumption is key to its identifiability results, as the non-Gaussianity enables the discovery of causal directions that remain ambiguous in purely Gaussian settings.
\end{itemize}
\BLACK

\subsubsection{Experimental rig}
  All tests were run on a laptop with an AMD Ryzen 9 5900HX, 32 GB of RAM, and an Nvidia RTX 3070 laptop GPU. The graphs were generated by interfacing with the pcalg package for R, which can be located on the CRAN website~\cite{PCALG1} (and pcalg implements the PC algorithm discussed above).  
  The release version of the package is pcalg version 2.7-9, and the specific function experimented on is the PC() function located in the pcalg package~\cite{PCALG2}\cite{PCALG1}.

\subsection{Results}

\subsubsection{RQ1: Do the projects demonstrate intra-project stability across releases?}\label{qrq1}

This first experiment applies PC to defect prediction.

As shown in Algorithm~\ref{alg1}, this experiment compares the graph from one project
between  release $i$ and release $i + 1$.   
 To implement this, we iterated over our datasets for each of our six projects, generating a new causal graph for each version. To keep everything consistent, we ran the ``stable" version of the PC function, with a full sample of each dataset, using the off-the-shelf defaults for
 our causal generator.  
 
  This algorithm is deterministic and will yield the same results every time it is run in this fashion. 
  
  This algorithm uses a default of $\alpha=0.01$ for the maximum probability of rejecting the null hypothesis. We use that value here since 
with the example code shown in the manual for this package,
$\alpha$ is always set to 0.01. In 
{\bf RQ3}, we experiment
with other values for $\alpha$.

\begin{algorithm}[h!tbp]
\footnotesize
	\caption{RQ1: Version Experiment Algorithm \newline\textbf{input: } d  datasets, used to generate and compare causal graphs.\newline \textbf{output: } a similarity report.}\label{alg1}
	\begin{algorithmic}[1]
		\For {($s \in \mathit{datasets}$)}
            \State $l \leftarrow newList()$
			\For {($v \in \mathit{s}$)}
				\State $m \leftarrow correlationMatrix(v)$
                \State $n \leftarrow numberOfRows(m)$
                \State $g \leftarrow pc(m, n, \alpha = .01)$
                \State $l.add(g)$
			\EndFor
            \State $exportAndSave(l)$
		\EndFor
    \State \textbf{Return} $compareSimilarity()$
	\end{algorithmic} 
\end{algorithm}

\begin{figure}[!t]
\caption{{\bf RQ1 results:} Graph of Jaccard Values across release, comparing release i to i + 1}\label{Version Instability}
\begin{center}
\includegraphics[width=8.5cm]{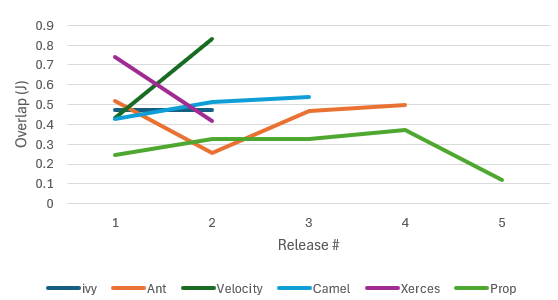}\end{center}\label{rq1}
\end{figure}

Fig.~\ref{rq1} shows the results. A slope between release $i$ and $i+1$ indicates that edges have changed between releases. Low y-axis values indicate poor overlap.

We note, in Fig.~\ref{rq1}, that in nearly all our observations, the Jaccard index seems to be under 0.5 on the y-axis; i.e., over half the causal connections found in one release are not seen in the next.
Hence, our answer to {\bf RQ1} is {\em that within one project across releases, there is much instability, despite running the supposedly ``stable'' version of the PC algorithm.  }

\subsubsection{RQ2: Do the graphs demonstrate inter-project stability across different projects?}
This second experiment applies PC to defect prediction data from different projects.
Here, Algorithm~2   checks if causal patterns hold in multiple projects. 

In order to observe this behavior we made the decision to use only the most recent datasets for all comparisons within this experiment.   The Prop dataset was excluded from this experiment as it contains differing variable counts which may impact graph generation beyond the intended scope of this experiment.

This experiment is deterministic and will yield the same results every time it is run in this fashion.
\begin{algorithm}[!t]
\footnotesize
    \caption{RQ2: Inter-Project Stability Experiment Algorithm \newline\textbf{input: }datasets,  being compared in this experiment\newline\textbf{output: }The similarity report for the causal graphs, exported and saved}
    \begin{algorithmic}[1]
        \State $graphList \leftarrow new List()$
        \For{(   $d  \in \mathit{datasets}$)}
            \State $m \leftarrow correlationMatrix(d)$
            \State $n \leftarrow numberOfRows(m)$
            \State $g \leftarrow pc(m, n, \alpha = .01)$
            \State $graphList.add(g)$
        \EndFor
        \For{($g \in \mathit{graphList}$)}
            \State $l \leftarrow newList()$
            \For{($i \in \mathit{graphList}$)}
                \If {$not g.equals(i)$}
                    \State $c \leftarrow compareGraphs(g, i)$
                    \State $l.add(c)$
                \EndIf
                
            \EndFor
            \State $exportAndSave(l)$
        \EndFor
    \end{algorithmic}
\end{algorithm}

\begin{table}[!t]
\caption{{\bf RQ2 results:} Inter-Project Jaccard Values. This matrix is symmetrical across the diagonal (so the values of blank cell $i,j$ can be found in
cell $j,i$).}
\label{tab:inter-project}
\begin{center}
\begin{tabular}{|l|l|l|l|l|l|} 
\hline
\footnotesize
\textbf{Project} & \textbf{Ivy} & \textbf{Ant} & \textbf{Velocity} & \textbf{Camel} & \textbf{Xerces}  \\ \hline 
\textbf{Ivy} & 1 & .28 & .37 & .25 & .26 \\ \hline
\textbf{Ant} &   & 1 & .28 & .19 & .20 \\ \hline
\textbf{Velocity} &   &  & 1 & .20 & .26 \\ \hline
\textbf{Camel} &   &   &   & 1 & .29 \\ \hline
\textbf{Xerces} &   &   &  &   & 1 \\ \hline

\end{tabular}
\end{center}
\end{table}

Fig.~\ref{tab:inter-project} shows the {\bf RQ2} results.
The Jaccard values indicate that around 75\% of the causal connections
found in one project are not present in the other
(exception: for  Ivy and Velocity, that overlap is 40\%). 
Hence, our answer to RQ2 is {\em that there exist wide, large  
 inter-project instabilities indicating very little consistency between graphs generated by the PC function on data from different projects.}

\subsubsection{RQ3: Does parameter tuning affect stability?}
This third experiment applied PC to the same defect prediction data sets, using different tuning parameters.
To explain this experiment: All learners come with ``magic'' tuning parameters that can
change and improve a learner's performance. Different data
sets can benefit from different tunings~\cite{fu2016tuning}. 

For example, the causality package used here is a tuning
parameter controlling the significance level or \(0 \le \alpha \le 1\), that sets a threshold for the conditional independence tests used in the PC function.  The lower the significance level, the lower the chance of a Type I error (or false-positive) occurring, and the higher it is, the higher the chance of a Type II error. This significance level also has a bearing on the total edge count of the graphs produced by the function, and the closer it gets to 1, the more edges the graph has.  Additionally, the smaller it gets, the number of edges in the graph appears to decrease all the way to very few or no edges.  

The default value for $\alpha$ in  our package is
$\alpha=0.01$, but researchers are free to change that value. For each data set, as shown in Algorithm~3,  we generated graphs from the latest version of the system, using  999 different 
$\alpha$ values ranging from  0.001 to 0.999.  These graphs were compared to the first graph generated using $\alpha=0.001$.

\begin{algorithm}[h!tbp]
\footnotesize
    \caption{RQ3: Parameter tuning   with significance level (\(\alpha\))\newline\textbf{inputs: }datasets,  used to generate the causal graph d\newline\textbf{outputs: } Similarity report, exported and saved}
        
    \begin{algorithmic}[1]
        \For{($d \in \mathit{datasets}$)}
            \State $graphList \leftarrow  newList()$
            \State $m \leftarrow correlationMatrix(d)$
            \State $n \leftarrow numberOfRows(m)$
            \For{$i \in {0...998}$}
                \State $sigLev \leftarrow (.001 + (i /1000))$
                \State $g \leftarrow pc(m, n, \alpha = sigLev)$
                \State $graphList.add(g)$
            \EndFor
            \State $g1 \leftarrow graphList.get(0)$ 
            \State $l \leftarrow newList()$
            \For{($i \in {1...998}$)}
                \State $g2 \leftarrow graphList.get(i)$
                \State $c \leftarrow compareGraphs(g1, g2)$
                \State $l.add(c)$
            \EndFor
            \State $exportAndSave(l)$
        \EndFor
    \end{algorithmic}
\end{algorithm}

Before reviewing the results, we state our pre-experimental expectations. Our prior work on hyperparameter optimization~\cite{fu2016tuning} showed that tuning can have small to medium effects on the veracity of the output (e.g., 10 to 30\% improvements in recall). These effects are often linear in the magnitude of the tuning. Hence, our
pre-experimental expectation was that changes to $\alpha$ would produce some, but perhaps not large, gradual changes to the causal
graph.

\begin{figure}[!t]
\caption{{\bf RQ3 results:} Graph of Jaccard values when shifting alpha \(\Delta = 0.001\). All graphs compared to the graph from $\alpha=0.001$.}\label{Variable tuning graph}
\begin{center}\includegraphics[width=8.5cm]{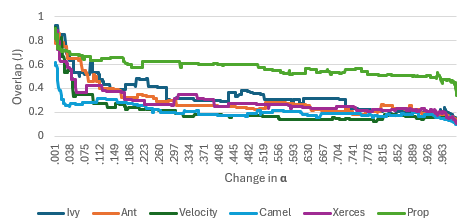}
\end{center}
\end{figure}

The {\bf RQ3} results are shown in the results. 
At larger $\alpha$ values (e.g. $\alpha > 0.2$), we can see the gradual and small changes suggested by our pre-experimental intuitions.
However, at lower $\alpha$ values (e.g. $\alpha < 0.1$), very small changes
in this parameter can lead to very large changes in the graphs (e.g., consider the range $0.025-0.075$ where many of the datasets exhibit precipitous drops).
 
The concern here is that brittleness in the resulting graphs is not documented
in the support material for this package. Hence, users of these tools might make a seemingly innocuous (and very small) change to $\alpha$ without realizing that the change can have a dramatic impact on the generated graph.


Aside:  one additional thing to note about this result is how the Prop dataset is less impacted by the significance level; we are unsure why this is due to the limitations of available datasets.  Our current hypothesis is that this may have to do with the size, as all of the other datasets are of similar sizes and exhibit similar end behaviors, while the prop is nearly one hundred times larger, making this seem like the most likely source of such a prominent disparity between the sets. 
 That said, Prop still exhibits the same behavior, just to a less extreme extent, as alpha decreases.  Other parameters could be modified and may potentially be sources of further instability, but just modifying the significance level displays a potential instability source.

Overall, our answer to {\bf RQ3} is {\em that tuning the parameters, specifically the significance level or \(\alpha\), can lead to significant instability in the graphs output by the generator.}

\subsubsection{RQ4: How much do causal graphs change if we remove only small fractions
     of the training data?}

     This fourth experiment  applies PC to the small variations of the same  defect prediction data sets. Here, we defined ``small'' to be ``remove 10\% of the data, selected at random''.
For each data set:
\begin{itemize}
\item This experiment ran 20 times with 90\% sub-samples of the data.
\item 19 of the graphs were compared to the first graph generated.
\end{itemize}
Before reviewing the results, we state our pre-experimental expectations: 
we expected that   removing  merely 10\% of the data should not have a large impact
on the generated edges.

\BLACK
 
\begin{algorithm}[h!tbp]
\BLACK
\footnotesize
    \caption{RQ4: Intra-Project 90\% Subsampling Algorithm\newline\textbf{inputs: }datasets  used to generate the causal graph d\newline\textbf{outputs: } Similarity report, exported and saved}
    \begin{algorithmic}[1]
         \For{($d \in \mathit{datasets}$)}
            \State $graphList \leftarrow newList()$
            \State $sampSize \leftarrow ceil(.9 * d.getRows())$
            \For{$i \in {0...19}$}
                \State $sample \leftarrow sample(d, sampSize)$
                \State $m \leftarrow correlationMatrix(sample)$
                \State $g \leftarrow pc(m, sampSize, \alpha = .01)$
                \State $graphList.add(g)$
            \EndFor
            \State $g1 \leftarrow graphList.get(0)$ 
            \State $l \leftarrow newList()$
            \For{($i \in {1...19}$)}
                \State $g2 \leftarrow graphList.get(i)$
                \State $c \leftarrow compareGraphs(g1, g2)$
                \State $l.add(c)$
            \EndFor
            \State $exportAndSave(l)$
         \EndFor
    \end{algorithmic}
\end{algorithm}

\begin{figure}[h]
\caption{{\bf RQ4 results:} Graph of 90\% subsample Jaccard values from datasets}\label{Subsample graphs}
\begin{center}
\includegraphics[width=8.5cm]{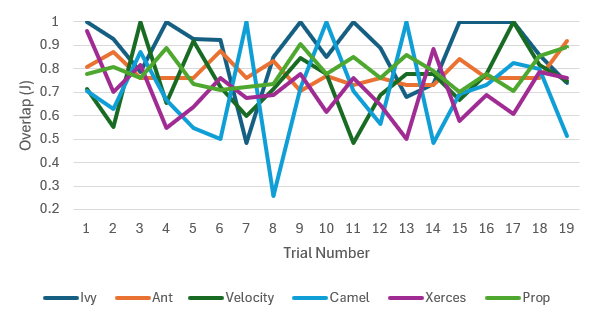}
\end{center}
\end{figure}

The results did not meet our expectations.
As seen in Fig.~\ref{Subsample graphs}, the Jaccard index between graphs can vary wildly between Jaccards of 0.6 to 1, with some observations slipping as low as 0.25.  As these sub-samples were generated randomly, there is a risk of obtaining certain outliers in the dataset, resulting in significantly different graphs, so more trials would be necessary to determine which datasets in particular are more susceptible to this behavior.  That said,   0.6 to 1 is still a fairly wide zone of variability.     

Overall, our answer to {\bf RQ4} is {\em that instability is not always present in a 90\% subsample, at least relatively compared to the other experiments.  In the subsample experiments, most of the Jaccard values were within the range of 0.6 to 1. However, this still represents a fair amount of variability across all of the datasets.}

\BLUE

\subsubsection{ RQ5: Do the  instabilities reported above hold for other causal generators and other tasks? }

\here{AE.1}  An issue with the above
analysis is that it is based on only
one causal graph generator (PC) and
only one SE task (defect prediction). 
To address that issue, this section explores our four research questions for  the three SE tasks described in \S\ref{Xdata} using the four causal graph generators described in \S\ref{algorithms};  i.e.

{ \footnotesize \[
\underbrace{\left(\begin{array}{@{\bf}c}
  RQ1\\ RQ2\\ RQ3\\ RQ4 
\end{array}\right)}_{\mathit{4\;research\;questions}}
\times\;\;\;\;
\underbrace{\left(\begin{array}{c}
PC\\
FCI\\
GES\\
LiNGAM
\end{array}\right)}_{4\;generators} 
\;\;\times\;\; 
\underbrace{\left(\begin{array}{c}
\mathit{18\;defect\; prediction}\\
\mathit{2 \;software\; configuration}\\
\mathit{3 \;software \;process\; management}
\end{array}\right)}_{\mathit{3\;tasks,\;23\; data\; sets}}
 \]}

\noindent
These experiments resulted in the same results seen above for   {\bf RQ1, RQ2, RQ3, and RQ4}. Hence, rather than repeat the same (uninformative)  charts and figures, we summarize the results in 
 Table~\ref{Revisionalternative alg performance comparisons}.  As shown in that table:
 \begin{itemize}
 \item  We restrict the {\bf RQ1} and {\bf RQ2} experiments to the defect data sets
 since, for that data, we can find multiple data sets with multiple releases, all
 with the same attributes.
  \item We restrict the {\bf RQ3}   changing $\alpha$ study to   PC and FCI since the other generators do not have that control parameter.
 \end{itemize}

\definecolor{worst}{HTML}{FFCCC9}
\definecolor{best}{HTML}{9AFF99}

 \begin{table}[]
\centering
\caption{ Jaccard indexes seen in different experiments. }\label{Revisionalternative alg performance comparisons}
\BLUE
\begin{tabular}{|l|l|p{1.2cm}|p{1.2cm}|p{1.2cm}|p{1.2cm}|}
\cline{3-6}
  \multicolumn{2}{c|}{~}& \multicolumn{4}{c|}{Causal graph generator}\\\hline
\textbf{data} & \textbf{question } & 
\textbf{PC} \newline $\mu$ $(\sigma)$
&  \textbf{FCI}\newline  $\mu$ $(\sigma)$
&  \textbf{GES} \newline  $\mu$ $(\sigma)$
& \textbf{LiNGAM} \newline  $\mu$ $(\sigma)$
\\ \hline
defect & {\bf RQ1} : across releases             & .67   (.07)                            & .43                     (.11)          & .50         (.08)                      & .39        (.06)    \\ \hline
defect & {\bf RQ2} : across projects          & .42        (.04)                      & .19                     (.07)          & .34          (.04)                     & .24         (0.5)     \\ \hline
defect & {\bf RQ3} : changing $\alpha$  &  .51 (.06)     &  .27   (.07)    & N/A                               & N/A     \\ 
config & {\bf RQ3} : changing $\alpha$                 & .27   (.07)                           & .10        (.04)                       & N/A                               & N/A           \\  
process & {\bf RQ3} : changing $\alpha$                &  .12  (.03)     &  .06  (.02)      & N/A                               & N/A    \\ \hline
defect & {\bf RQ4} : subsets            &  .84  (.06)     &  .59  (.12)     & .70         (.07)                      & .52                 (.10) \\ 
config & {\bf RQ4} : subsets            & .75   (.18)                            & .59  (.26)     &  .73  (.16)     &  .57  (.20)   \\ 
process & {\bf RQ4} : subsets           &  .64  (.11)     &  .54  (.13)     &  .59  (.09)     &  .29  (.07) \\\hline 
\end{tabular}
\BLACK
\end{table}

\noindent
Two salient features of   Table~\ref{Revisionalternative alg performance comparisons} are:
\begin{itemize}
\item As commented above, causal graph instability is large for defect prediction.
When we change treatments:
\begin{itemize}
\item
At least a third of the edges disappear for  $\frac{11}{14} $of those results;
\item
At least half the edges disappear for $\frac{9}{14}$ of those results;
\end{itemize}
\item While instability is poor in defect prediction, it is even worse in other tasks.
The results from {\bf RQ3} and {\bf RQ4} offer six cells of Table~\ref{Revisionalternative alg performance comparisons}, where we can compare results from defects, config, and process. In $\frac{4}{6}$ of those cells, the mean Jaccard indexes are smaller (i.e., worse)  in config and process.  
\end{itemize}

We were curious if particular generators, tasks, or experiments led to worse instabilities.
To address that question, we employed the  Scott-Knott
procedure of Table~\ref{sk}. This Scott-Knott procedure employs non-parametric statistics (bootstrap~\cite{Efron:526679}
and effect size~\cite{macbeth2011cliff,hess2004robust}) to cluster together similar results.
As shown in Table~\ref{Stat_analysis}, this procedure sorts treatments by their mean. In that sort, 
   item $i+1$   gets a new rank if it is statistically distinguishable by more than a
small effect from item $i$.

\begin{table}[!t]
\caption{ Details on the Scott-Knott procedure for ranking sorted treatments. }\label{sk}
\BLUE
\begin{tabular}{|p{.95\linewidth}|}\hline
To rank all our experiments, we use a Scott-Knott analysis~\cite{mittas2012ranking}.  
 Procedurally, Scott-Knott is a top-down clustering procedure that recursively splits a list of treatments
at that treatment that   maximizes the difference $E(\Delta)$ in the expected mean value before and after the split:
\begin{equation}
    E(\Delta) = \frac{|l_1|}{|l|}abs(\overline{l_1} - \overline{l})^2 + \frac{|l_2|}{|l|}abs(\overline{l_2} - \overline{l})^2
\end{equation}
where:
\begin{itemize}
    \item
    $|l|$, $|l_1|$, and $|l_2|$: Size of list $l$, $l_1$, and $l_2$.
    \item
    $\overline{l}$, $\overline{l_1}$, and $\overline{l_2}$: Mean value of list $l$, $l_1$, and $l_2$.
\end{itemize}
Recursion continues if the two parts of the split are
statistically distinguishable by more than a
small effect 
A non-parametric bootstrap is used to test whether two populations are statistically distinguishable~\cite{Efron:526679}. 
Cliff's delta~\cite{macbeth2011cliff,hess2004robust} checks for small effects. Given  two lists $A$ and $B$ of size $|A|$ and $|B|$ then two
lists are different by a small effect if:
\[
\left(
\begin{array}{ c}
    \frac{\sum\limits_{x \in A} \sum\limits_{y \in B} \left\{ \begin{array}{l} +1, \mbox{  if $x > y$}\\
                    -1, \mbox{   if $x < y$}\\
                     0,  \mbox{   if $x = y$}
                \end{array} \right.}{|A||B|}
                \end{array}
                \right) \ge  0.147
 \]    \\\hline           
\end{tabular}
\BLACK
\end{table}
The salient features of   Table~\ref{Stat_analysis} are:
\begin{itemize}
\item
Median instabilities are .53; i.e., usually, nearly half the edges
found in one treatment can disappear in another; 
\item
Changing $\alpha$  produces most instability;
\item
Exploring subsets produces the least instability;
\item As to what causal graph generator was most unstable,
FCI was usually found in the lower half of this list.
\end{itemize}
\begin{table} 
\centering
\caption{Statistical Analysis using Cliff's Delta and Bootstrap with Skott-Knott for rank division.}
\label{Stat_analysis}
\scriptsize
\BLUE
\begin{tabular}{|l|l|l|c|c|}
\hline
Rank & Dataset & RQ and experiment & Alg & Mean (Stdev) \\ \hline
0    & defect     & {\bf R4} : subset            & PC      & .85 (.05)  \\ \hline
1    & config     & {\bf R4} : subset            & PC      & .73 (.28)  \\
     & config     & {\bf R4} : subset            & GES     & .73 (.25)  \\ \hline
2    & defect     & {\bf R4} : subset            & GES     & .70 (.07)  \\ \hline
3    & defect     & {\bf R1} : across releases   & PC      & .69 (.07)  \\ \hline
4    & proceess   & {\bf R4} : subset            & PC      & .64 (.09)  \\ \hline
5    & proceess   & {\bf R4} : subset            & GES     & .59 (.09)  \\
     & defect     & {\bf R4} : subset            & FCI     & .59 (.11)  \\ \hline
6    & proceess   & {\bf R4} : subset            & FCI     & .55 (.10)  \\
     & config     & {\bf R4} : subset            & LiNGAM  & .54 (.27)  \\
     & config     & {\bf R4} : subset            & FCI     & .53 (.22)  \\
     & defect     & {\bf R1} : across releases   & GES     & .53 (.06)  \\
     & defect     & {\bf R4} : subset            & LiNGAM  & .52 (.11)  \\
     & defect  & {\bf R3} : changing $\alpha$                        & PC      & .51 (.06)  \\ \hline
7    & defect     & {\bf R1} : across releases   & FCI     & .49 (.16)  \\ \hline
8    & defect  & {\bf R2} : across projects & PC      & .43 (.07)  \\ \hline
9    & defect     & {\bf R1} : across releases   & LiNGAM  & .41 (.06)  \\ \hline
10   & defect  & {\bf R2} : across projects & GES     & .35 (.05)  \\ \hline
11   & proceess   & {\bf R4} : subset            & LiNGAM  & .28 (.04)  \\
     & defect     & {\bf R3} : changing $\alpha$ & FCI     & .27 (.08)  \\
     & config     & {\bf R3} : changing $\alpha$ & PC      & .26 (.10)  \\
     & defect  & {\bf R2} : across projects & LiNGAM  & .25 (.05)  \\ \hline
12   & defect & {\bf R2} : across projects & FCI     & .22 (.14)  \\ \hline
13   & proceess   & {\bf R3} : changing $\alpha$ & PC      & .12 (.04)  \\ \hline
14   & config     & {\bf R3} : changing $\alpha$ & FCI     & .10 (.05)  \\ \hline
15   & proceess   & {\bf R3} : changing $\alpha$ & FCI     & .06 (.03)  \\ \hline
\end{tabular}
\BLACK
\end{table}

Overall, our answer to {\bf RQ5} is that  
{\em the instabilities reported above hold for other causal generators
and other tasks}.

\BLACK

{\BLUE}
\clearpage\section{Discussion}

\BLUE
\here{R1.0}
The size of the instabilities reported here is so large that it leads to the obvious question: {\em ``how to improve causal graph generation?''}.

Based on this work, we suspect that different causal graph generators are most appropriate for different kinds of data.  Elsewhere, we are exploring:
\begin{itemize}
\item
Listing the assumptions of each generator 
\item Using those lists to match between a particular generator and a particular data set. 
\end{itemize}
Preliminary results suggest that this approach can partially reduce instability.  We do not present that work here since  (a)~ they are still only very preliminary; and (b)~before the research community accepts a solution to a problem, first that problem must be documented (hence this paper).
 
\BLACK
  
\section{Threats to Validity}
Any study is challenged by threats to validity.

\textbf{Internal validity}  is the degree of confidence that the causal relationship you are testing is not influenced by other factors or variables.
We have tried a range of factors that might cause instability (intra vs. inter project changes, a hyperparameter change, some sub-sampling), and large instability was seen in all cases. 
 
\textbf{Sampling bias  } refers to instabilities in the conclusion due to
the data used in the study. No data mining study can explore all data sets, and it is possible that the instabilities seen here are some quirk of defect prediction data. In terms of future work, it would be useful to determine if different kinds of data generate more/less stable causal graphs.

\textbf{Construct validity} 
 is about how well a test measures the concept it was designed to evaluate.
 Our measure of instability is the Jaccard index, which looks at the internal edges of a graph. An alternate measure might be to ignore the internals and ask what inputs usually lead to what outputs. That said, we know of no publication using that measure.

\textbf{Dependability:} 
Non-deterministic experiments introduce a threat to validity (since conclusions may be a quirk of the random number generator). Most of our experiments are deterministic, with the exception of {\bf RQ4}.
To mitigate this threat to validity, we ran {\bf RQ4} 20 times with different random number seeds.

\section{Future Work}\label{why}

We hope this paper inspires much subsequent research. For example:
\begin{itemize}
 


\item
    
    \textbf{What is the impact of instability on the use cases of causal graphs?} How could the instability detected in this study impact the different ways people use causal graphs? For example, when diagnosing faults, is causal graph instability actually a good thing (since it generates more options)? Or, can the same final answers be obtained through different means across different graphs?   
    

   \item 

    \textbf{Beyond the Jaccard index, are there better ways to evaluate the structure of a causal graph?}  If so, how can this value be optimized, and is this a metric that can be consistently applied and evaluated in graphs automatically?  Additionally, would it be possible to use an evaluation function for a graph generator to promote stability? For example, if a causal relationship is known before running the generator, can this evaluation function be asserted so that the generator begins with a partial graph? And would this promote stability and/or utility between graphs?  
    

 
\end{itemize} 
 More ambitiously, we ask
 \textbf{how to mitigate for instability?} Do stochastic approaches improve stability? Does lowering the significance level when testing for conditional independence promote stability, as it decreases false positives and reduces the overall total edge count? If the decreasing significance level does promote stability, what is a desirable significance level to choose so the graph can still be useful while also being consistent? Additionally, can the correct tuning parameters be automatically selected in order to tune them properly for a given dataset to make the best guess at what the correct configuration would be, if one exists, to generate reliable causal graphs for that dataset?

We conjecture that there are many possible reasons for this instability, and future work should address those possibilities.  
One root cause for instability that would be hard to address
is the inherent complexity of causal computation.
As mentioned above, causal graph generation is NP-hard. Hence,   a practical and scalable causal graph generator must use heuristics to generate its models.
By their very definition, such heuristics are incomplete, so seemingly minor changes in the control parameters of the generator or in the data sampling can lead to very different graphs. Perhaps to mitigate for instability, we need a simpler definition of causality that is not so problematic.

 One major issue is that causal graphs typically assume a static environment where relationships between variables remain constant over time~\cite{10.5555/2074284.2074334}. However, in real-world software systems, these relationships can change due to updates, new features, or changing user behavior. For example, a performance optimization that works under current conditions might become ineffective or even detrimental after a software update or a change in user traffic patterns. 
  In highly dynamic environments, the causal relationships between variables can shift, leading to instabilities in the causal graph~\cite{dong2014modeling}. For instance, in a distributed system, network latency might suddenly increase due to changes in infrastructure or varying loads. If the causal graph does not account for such variations, it may fail to accurately predict performance issues or the impact of changes. To mitigate this, engineers need to continuously monitor the system and update the causal graph to reflect current conditions, ensuring that causal reasoning remains accurate and relevant.

 Incomplete data can also lead to instabilities in causal graphs~\cite{rohrer2018thinking, 10.5555/2073796.2073813}. When certain variables or causal relationships are not included, the graph provides an incomplete picture, potentially leading to incorrect causal inferences. For example, in a software application, if user behavior data is not fully captured, the causal graph may miss critical relationships that impact system performance or user experience. Engineers need to ensure that they have comprehensive data and regularly update the causal graph to include new variables and relationships as they become known.

As software systems evolve, their requirements and the interactions between components often change~\cite{9218193}. This evolution can introduce new variables and alter existing causal relationships, leading to instabilities in the causal graph. For example, adding a new feature might introduce dependencies that were not previously considered, changing the overall system dynamics. Engineers must be proactive in updating the causal graph to reflect these changes, ensuring that it remains an accurate representation of the system's causal structure.
  
\section{Conclusion}
Our reading of the literature indicates that causality is a topic of growing interest in software engineering. But something can be widely used and still be wildly misleading.
The experiments of this paper have
demonstrated  causal graph instability:  
\begin{itemize}
\item Across different releases of the same project (see {\bf RQ1});
\item Across different projects (see {\bf RQ2});
\item Across tuning changes to the data of one project (see {\bf RQ3});
\item Across very small changes (10\%) of the training data (see {\bf RQ4}).
\item  \textcolor{black}{These
effects were observed
in three different SE domains and for four different causal graph generators (see {\bf RQ5}).} 
\end{itemize}
From these experiments and our literature review, we make five observations:
\begin{enumerate}

 \item
Many SE researchers do not appreciate the frequency and impact of causal instability.
\item
Generalizations based on causal graphs now have an extra threat to validity, i.e., that their results may not hold across even minor variations to the training methods. 
\item
Before using causal graphs  to make   general claims about software development,  SE researchers need to test   that any such conclusion 
 {\em holds over the numerous causal graphs} that we might generate from the same data.
 \item
Unless the second point is satisfied,  we should not use causal graphs as a way to generate the models
that are reported to business users or the research community.
\item 
As recommended by some statisticians, perhaps it is best to use causal graphs as a verification tool.  
 Siebert~\cite{SIEBERT2023107198} argues that 
{\em 
``The main purpose of such a (causal) graph is to reason about the presence of potential spurious correlations
that may bias the analysis''}.
In this view, causal graphs are an interim, possibly throw-away, model that is generated and inspected before moving on, perhaps to some other modeling methods
\end{enumerate} 
As an example of the last point,  based on our background knowledge of software engineering, we suspect that there is something very wrong about 
 the data used to create Fig.~\ref{cause1}.a.  Lines of code should  {\em cause} bugs (and not the other way around).  Hence, before doing anything else with the Fig.~\ref{cause1} data, there needs to be some investigation into that anomaly.

\newpage
\section*{Declarations}

\begin{description}
\item[\bf Funding: ] \hfill \\
The authors have no relevant financial or non-financial interests to disclose.
All authors certify that they have no affiliations with or involvement in any organization or entity with any financial interest or non-financial interest in the subject matter or materials discussed in this manuscript.
No funding bodies were involved in the creation
of this work.
\item[\bf Ethical approval:] \hfill \\
Lacking human or animal subjects,
an ethical review was not required. 
\item[\bf Informed consent:] \hfill \\
This study had no human subjects.
\item[\bf Author Contributions:] \hfill \\
All this paper's experimental work was conducted by Jeremy Hulse.
Jeremy Hulse and Nasir Eisty and Tim Menzies contributed equally
to the writing of this paper. 
\item[\bf Data Availability Statement:] \hfill \\
To repeat and/or refine and/or refute this work, see our scripts and data at
\url{https://github.com/jphulse/Stability_Of_Causal_Graphs_Public}.
\item[\bf Conflict of Interest:] \hfill \\ The authors declared that they have no conflict of interest.
The authors have no competing interests to declare that are relevant to the content of this article.
\item[\bf Clinical Trial Number:] \hfill \\
Clinical trial number: not applicable.
\end{description}

\bibliographystyle{acm}
\bibliography{bibliography,bibliography_NE}
\section*{Biography and Photo}

\vspace{1em}
\noindent
\begin{minipage}{0.3\textwidth}
    \includegraphics[width=\textwidth]{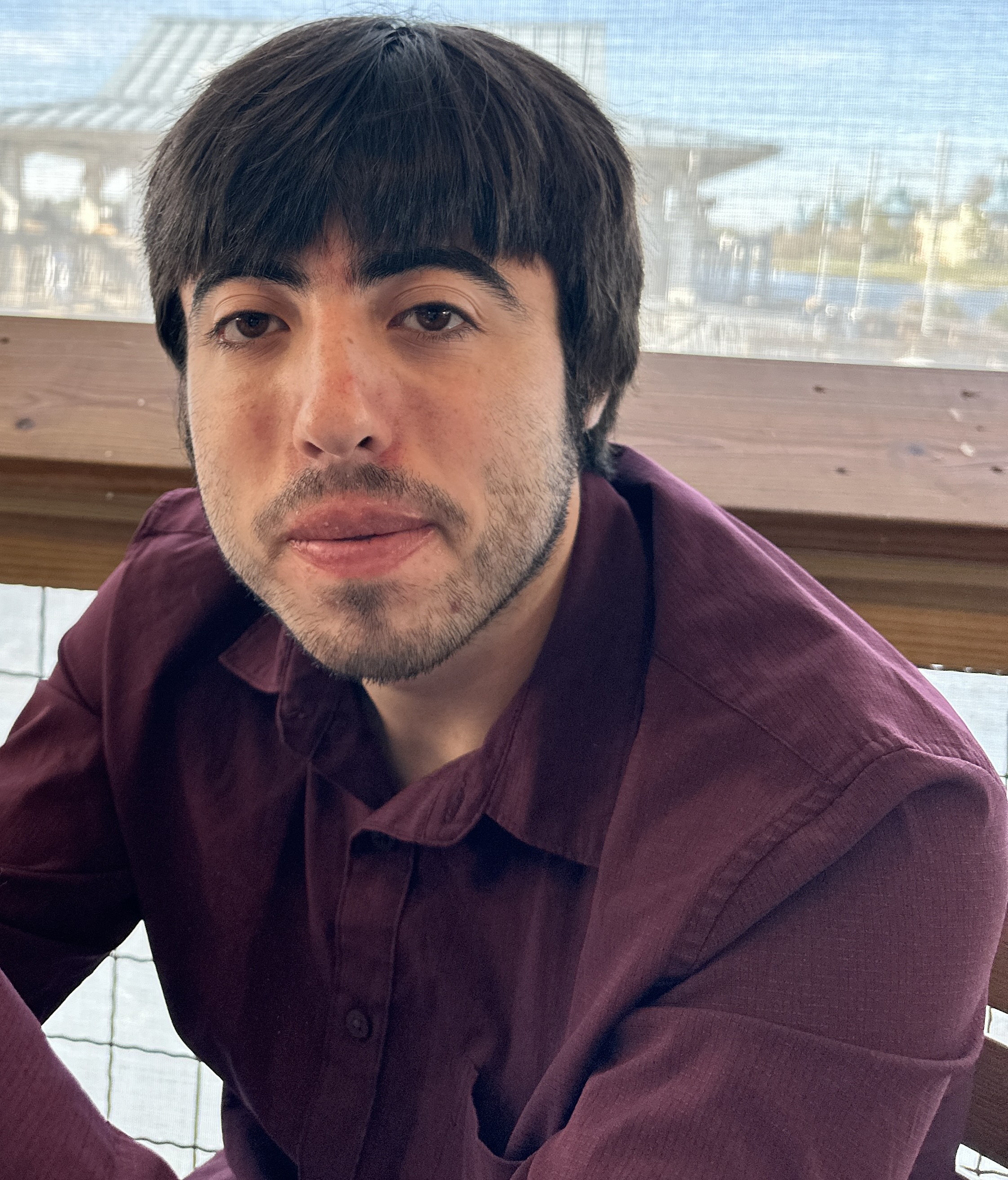} 
\end{minipage}
\hfill
\begin{minipage}{0.65\textwidth}
         \textbf{Jeremy Hulse} is a student and emerging researcher pursuing a Master's degree in Computer Science at North Carolina State University.  He aims to explore procedural generation, VR/AR technologies and software, and the logical representations of dynamic and interactive software systems, with a focus on building expertise and contributing to advancements in these areas. He received his bachelor's degree in Computer Science from North Carolina State University in 2025. Contact him at jphulse@ncsu.edu.
\end{minipage}
\vspace{1em}
\noindent
\begin{minipage}{0.3\textwidth}
    \includegraphics[width=\textwidth]{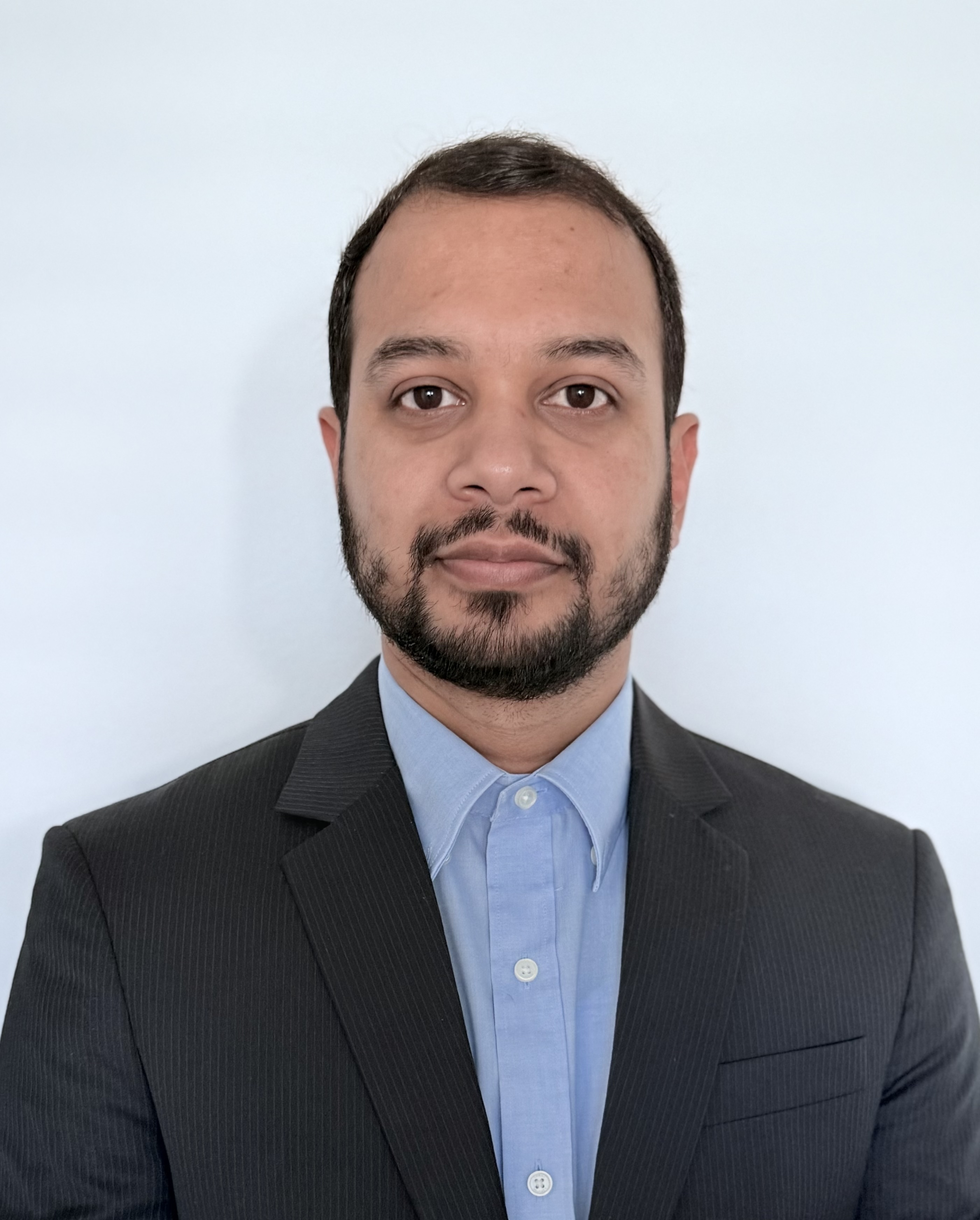} 
\end{minipage}
\hfill
\begin{minipage}{0.65\textwidth}
         \textbf{Nasir U. Eisty} is an Assistant Professor of Computer Science at the University of Tennessee Knoxville. His research interests lie in the areas of Empirical Software Engineering, AI for Software Engineering, Scientific and Research Software Engineering, and Software Security. He received his Ph.D. degree in Computer Science from the University of Alabama. Contact him at neisty@utk.edu
\end{minipage}

\vspace{1em}
\noindent
\begin{minipage}{0.3\textwidth}
    \includegraphics[width=\textwidth]{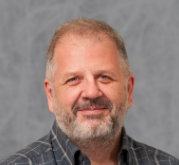} 
\end{minipage}
\hfill
\begin{minipage}{0.65\textwidth}
         \textbf {Tim Menzies} (ACM Fellow, IEEE Fellow, ASE Fellow, Ph.D., UNSW, 1995) is a full Professor in Computer Science at North Carolina State. He is the director of the   Irrational Research Lab (mad scientists r'us) and the author of over 300 publications (refereed) with 24,000 citations and an h-index of 74. He has graduated 22 Ph.D. students, and has been a lead researcher on projects for NSF, NIJ, DoD, NASA, USDA, and private companies (total funding of \$19+ million). Prof. Menzies is the editor-in-chief of the Automated Software Engineering journal and associate editor of TSE and other leading SE journals. For more, see  \url{https://timm.fyi}.
\end{minipage}


\end{document}